\documentstyle[epsf]{mn}

\title{The discovery of very red counterparts to faint X-ray sources}
\author[A. Newsam, I. M$^{c}$Hardy,  L. Jones and K. Mason]{
  A.M.~Newsam,$^1$ I.M.~M$^{\rm c}$Hardy,$^1$ L.R.~Jones,$^{2,3}$
	K.O.~Mason$^4$\\
  $^1$Department of Physics and Astronomy, Southampton University, 
	Southampton, SO17 1BJ.\\
  $^2$School of Physics and Space Research, University of Birmingham,
	Edgbaston, Birmingham B15 2TT.\\
  $^3$Code 660, NASA/GSFC, Greenbelt, MD 20771, USA.\\
  $^4$Mullard Space Science Laboratory, University College London, 
	Holmbury St Mary, Dorking RH5 6NT.}
\date{Submitted to MNRAS March 1997, Accepted August 1997}

\def\etal{\em et~al.}
\def\rosat{\sc Rosat}
\def\lxlopt{${\rm L}_{\rm X}/{\rm L}_{\rm opt}$}
\def\lglxlopt{$\log[{\rm L}_{\rm X}/{\rm L}_{\rm opt}$]}

\def\scite#1{\def\citename##1##2{##1}\csname b@#1\endcsname
 \shortcite{#1}}
\def\pcite#1{\def\citename##1##2{##1##2}\csname b@#1\endcsname\nocite{#1}}
\def\ycite#1{\def\citename##1##2{##2}\csname b@#1\endcsname\nocite{#1}}
\def\ncite#1{\def\citename##1##2{##1}\csname b@#1\endcsname\nocite{#1}}

\begin{document}

\maketitle

\begin{abstract}

We present deep K-band imaging at the positions of four very faint X-ray
sources found in the UK {\rosat} Deep Survey \cite{mn_nov} to have no optical
counterpart brighter than R$\sim$23. Likely identifications are found within
the {\rosat} error circle in all four fields with R$-$K colours of between
3.2$\pm$0.4 and 6.4$\pm$0.6. From a consideration of the R$-$K colours and
X-ray to optical luminosity ratios of the candidate identifications, we
tentatively classify two of the X-ray sources as very distant ($z\sim 1$)
clusters of galaxies, one as a narrow emission line galaxy and one as an
obscured QSO.

\end{abstract}


\section{Introduction}

Optical identification of sources detected in deep X-ray images, particularly
from the {\rosat} satellite, have, in recent years, proved a highly successful
probe of the origins of the soft Cosmic X-ray background (XRB)
\cite{Shanks+91,Boyle+94,mn_nov}. In particular, it has been seen that at
bright fluxes, the major class of contributors is QSOs. However, they are
unlikely to contribute more than $\sim$50\% of the X-ray background at 1keV
\cite{Boyle+94,Jones+96,mn_nov} and, even were their contribution to increase
greatly beyond the limits of current surveys, their characteristic X-ray
spectrum is too soft to match the residual XRB \cite{R-C+96}. Deeper X-ray
surveys, however, are showing that at fainter fluxes, there is a new class of
galaxies with narrow optical emission lines (NELGs) which have suitably hard
X-ray spectra \cite{McHardy+95/6,R-C+96,Almaini+96}.

The deepest X-ray survey yet optically identified is the UK {\rosat} Deep
Survey. This reaches a depth of $2 \times 10^{-15}$ erg cm$^{-2}$ s$^{-1}$
(0.5-2keV) and is described in detail in \scite{mn_nov}.

Within the ``complete'' sample area defined by \scite{mn_nov}, there are eleven
unidentified sources (out of a total of 70 sources). Of the unidentified
sources, most (7) remain unidentified because adequate optical spectra have not
been obtained for all likely candidates. However, for four of the X-ray sources
there are no optical candidates brighter than ${\rm R}\sim 23$. This is a
puzzle since they are not primarily the faintest X-ray objects in the
survey. One possibility is that the source of the X-rays are obscured
AGN~--~long proposed as a contributor to the XRB (see for example
\pcite{Shanks+96}). The discovery of such objects would have important
implications for the origin of the hard X-ray background. The X-rays could also
be coming from distant clusters of galaxies, which can have high X-ray to
optical luminosity ratios (see, for example, \pcite{Stocke+91}). Other
possibilities include further NELG objects or very high redshift QSOs, where
the bright Ly$\alpha$ emission has been redshifted out of the R-band. All of
these classes of objects, particularly the obscured AGN, will have red
optical/infra-red colours, so in 1996 we performed deep K-band imaging at UKIRT
of the four `blank' Deep Survey fields to identify candidate objects and try to
determine their nature. The results of this imaging are presented in this
paper.

In section~\ref{sec:survey} we briefly describe the X-ray and optical data that
make up the deep survey, in section~\ref{sec:irdata} we present the K-band
images and in~\ref{sec:ident}, we consider possible scenarios.
Finally in section~\ref{sec:conclusions} we discuss the possible significance
of the results.

\section{The UK ROSAT Deep Survey: Optical Identifications}%
\label{sec:survey}

The UK {\rosat} Deep Survey is a total of 115 ksec of {\rosat} position
sensitive proportional counter (PSPC) observation of RA~$13\,\, 34\,\, 37.0$
Dec~$+37\,\, 54\,\, 44$ (J2000) divided between two exposures~---~in June 1991
\cite{GBR} and June/July 1993. This region of sky was selected because of its
extremely low obscuration~---~$N_H \sim 6.5 \times 10^{19}
\hbox{cm}^{-2}$. Using these data, sources were identified at more than
3.5$\sigma$ significance down to $1.6 \times 10^{-15} \hbox{ erg cm$^{-2}$
s$^{-1}$}$. Optical CCD images of the inner 15 arcmin radius of the {\sc Rosat}
field taken at the University of Hawaii 88inch telescope were then used to find
potential optical counterparts and subsequent spectroscopy has enabled
identification of the sources in a statistically complete subsection making up
about 80\% of the 15 arcmin survey area down to $2 \times 10^{-15} \hbox{ erg
cm$^{-2}$ s$^{-1}$}$ \cite{mn_nov}.

The majority of these sources are identified with QSOs, NELGs and
galaxy clusters, and a small number with stars. However, of interest
in this paper are four X-ray sources where the initial CCD imaging
revealed no candidates within $\sim 15$ arcsec of the X-ray position
brighter than $\rm{R} \sim 23$. All four of these sources are towards
the fainter flux end of this sample (ie $< 5 \times 10^{-15} \hbox{
erg cm$^{-2}$ s$^{-1}$}$).
%
%
\begin{figure}
 \epsfxsize \hsize
 \epsffile{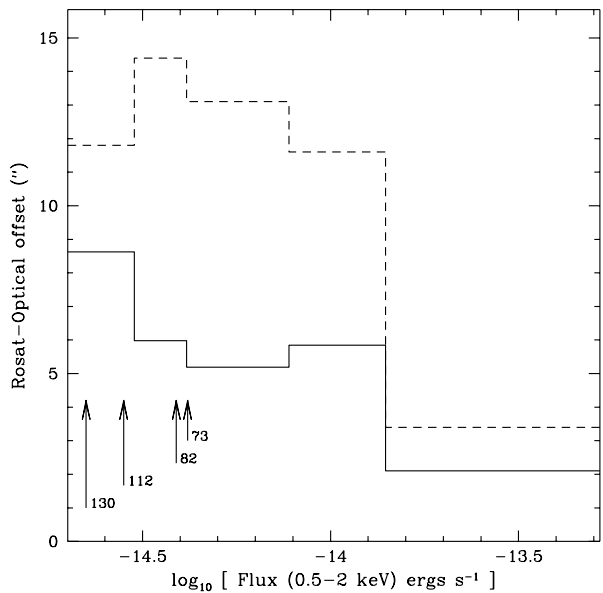}
 \caption{The offset between {\rosat} positions and optical
  counterparts as a function of X-ray flux. The average offset in each
  flux bin is shown by the solid line, and the maximum offset by the
  dashed line.  Only firm identifications are included, and galaxy
  clusters and groups are excluded because of the uncertainty that
  their extension introduces into the X-ray position.  The flux bins
  are chosen to each contain a similar number of objects. The flux
  positions of the four blank field sources are shown.
  }\label{fig:offsets}
\end{figure}
However, even down at these faint limits, we have found from the offsets
between the optical and X-ray positions of identified sources as shown in
figure~\ref{fig:offsets}, that we would still expect the optical counterpart of
the faintest of these sources to be within $\sim 12$ arcsec of the {\rosat}
position (90\% confidence).

The source searching in the deep survey was based on the Cash statistic
\cite{Cash79} and is described in more detail in \scite{mn_nov} and
\scite{GBR}. Simulations showed that for sources detected at greater than
3.5$\sigma$ significance, the survey is complete above a flux limit of 1.8
$\times$ 10$^{-15}$ erg cm$^{-2}$ s$^{-1}$. Therefore, even the faintest of the
sources (r130~--~see table~\ref{tab:candidates}) is detected at $>$
3.5$\sigma$.

In March 1995, the whole 15 arcmin radius {\rosat} survey region was
again imaged in the R band to a greater depth, using the University of
Hawaii 8K$\times$8K CCD array \cite{8kpaper} on CFHT with a 1 hour
exposure. This image revealed some possible optical candidates in two
of the fields at R$\sim$23, but nothing of note (ie R$\la$24) in the
other two (see the right hand panels of figures~\ref{fig:r73_piccies}
to~\ref{fig:r130_piccies}).

\section{Infra-red Imaging}\label{sec:irdata}

The K-band images were obtained in one night in May 1996 at UKIRT using the
IRCAM3 camera. The images in figures~\ref{fig:r73_piccies}
through~\ref{fig:r130_piccies} were formed from multiple 9-point
mosaics~---~hence the loss of quality towards the edges where fewer of the
sub-images overlap. Total exposure times range from just over 20 mins to 1 hour
(see table~\ref{tab:kimage}). Initial reduction was performed at the telescope
using the {\sc ircam3\_clred} package, but the final images given here were
re-reduced in IRAF using purpose-built codes.  The seeing for the night was
just greater than 1 arcsec and the pixels are 0.286 arcsec (the original K-band
IRCAM3 images were 256$\times$256 pixels, the mosaics created are
311$\times$311 pixels).

The four K-band images are given in the left hand panels of
figures~\ref{fig:r73_piccies} to~\ref{fig:r130_piccies}, the right hand panels
being the deepest available R-band images of the same areas for comparison. It
should be noted that exposure times vary from image to image, and the
grey-scale shading of each has been chosen individually for each image for
clarity. Direct visual comparisons {\em between\/} the different K-band images
should not, therefore, be made.

\begin{table}
\begin{tabular}{cccc}
Object &  Total Exposure & Figure & Limiting \\
 Name  &    (secs)       & Number & K mag    \vspace{1mm}\\
  r73 & 1620 & \ref{fig:r73_piccies}  & $\sim$20  \\
  r82 & 1620 & \ref{fig:r82_piccies}  & $\sim$20  \\
 r112 & 3240 & \ref{fig:r112_piccies} & $\sim$20.5\\
 r130 & 1320 & \ref{fig:r130_piccies} & $\sim$20  \\
\end{tabular}
\caption{The details of the K-band images in figures~\ref{fig:r73_piccies}
 to~\ref{fig:r130_piccies}. The limiting magnitudes refer to the central
140$\times$140 pixel square region where all the images in the 9-position
sampling pattern overlap.}\label{tab:kimage}
\end{table}
We will now look at the results for each of the fields
individually. Table~\ref{tab:candidates} gives the relevant X-ray and
optical/IR details of the objects. The magnitudes given refer to the
candidates discussed in the sections below.
\begin{figure}
 \epsfxsize \hsize
 \epsffile{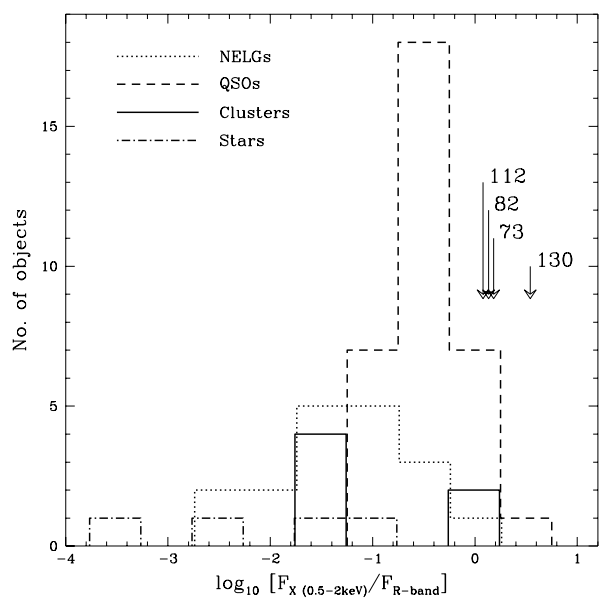}
 \caption{A comparison of the distribution of X-ray to optical flux ratios for 
  the firmly identified QSOs, NELGs and stars in the Deep Survey and the 
  ratios for the best candidates found in the blank fields. The histograms show
  the distributions of the three classes of known objects, with the positions
  of the four blank field candidates marked by arrows.
  The ${\rm F}_{\rm X}$ is the integrated $0.5$--2 keV {\rosat} flux, 
  and ${\rm F}_{\hbox{\scriptsize R-band}}$ the optical R band flux.
}\label{fig:fxfopt}
\end{figure}

\subsection{Object r73}\label{sec:r73}

\begin{figure*}
 \begin{center}
  \leavevmode
  \epsfxsize 0.9\hsize
  \epsffile{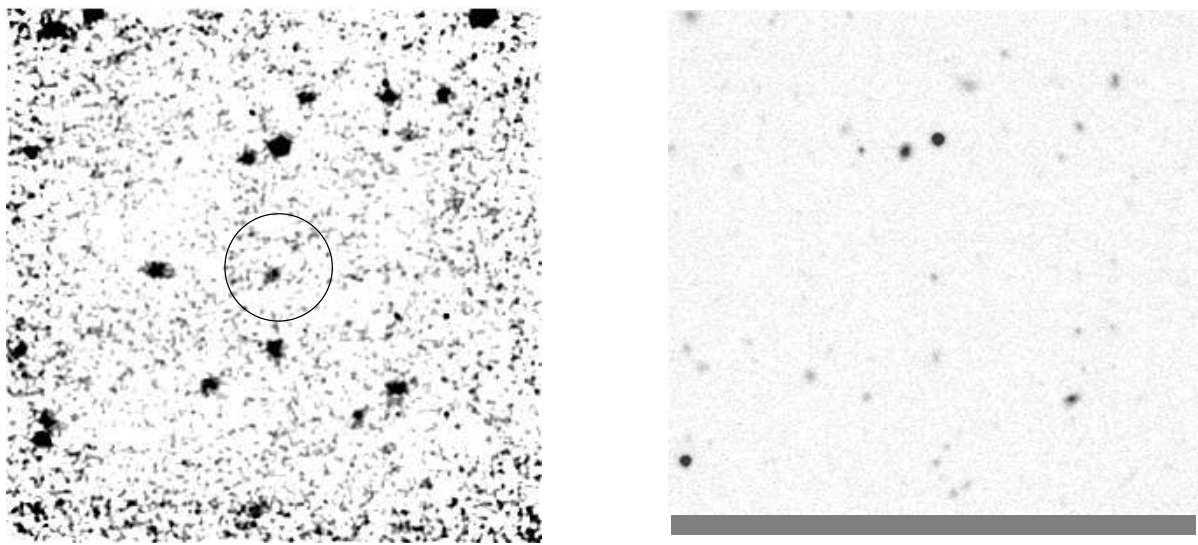}
 \end{center}
 \caption{Object r73. In the left hand panel is shown the K-band image of 
  object r73. The image has been smoothed slightly for clarity (with a 
  gaussian with $1\sigma = 1$ pixel). A 10 arcsec radius error circle around 
  the approximate {\rosat} position is given and the object near the center is 
  the one referred to in the text. The right hand panel has an R-band image
  for comparison.
  For both images, the north is to the top and east to the left. The K-band
  image has a limiting magnitude of K$\sim$20 and the R-band image, R$\sim$24.
}\label{fig:r73_piccies}
\end{figure*}
Object r73 is the X-ray brightest of the four sources (see
table~\ref{tab:candidates}) and a very faint (R$\sim$23) object can be seen
close to the X-ray centroid in our deepest optical images (this can just be
seen at the centre of the right hand frame of
figure~\ref{fig:r73_piccies}). The only other reasonable candidates are the
pair of objects to the north (up) which are 20 and 21mag. However, these are
both 20 arcsec from the X-ray centroid, which is far larger than the average
positional offsets at this flux (cf figure~\ref{fig:offsets}).

The K-band image shows no additional candidates close to the X-ray
centroid. A faint, red object (K$\sim$18.6, R$-$K$\ga$5) is visible to
the east, but is again, nearly 20 arcsec from the centroid, so the
central source is by far the most likely candidate. The relevant
magnitudes for this central object are given in
table~\ref{tab:candidates}.

\subsection{Object r82}\label{sec:r82}

\begin{figure*}
 \begin{center}
  \leavevmode
  \epsfxsize 0.9\hsize
  \epsffile{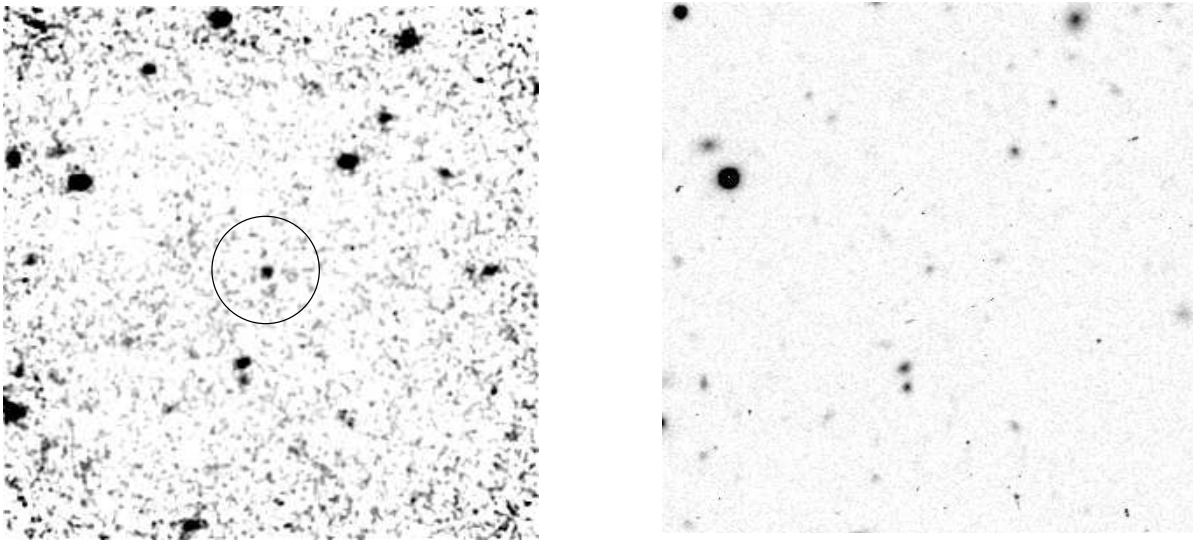}
 \end{center}
 \caption{Object r82. As figure~\protect\ref{fig:r73_piccies}.
}\label{fig:r82_piccies}
\end{figure*}
For object r82 there is a clear-cut candidate at the center of the
X-ray error circle (marked in the left hand panel of
figure~\ref{fig:r82_piccies}) which is also just visible on the deep
R-band image. The faintness of the candidate identifications in all of
the K-band images prevents the reliable detection of extension, but
the central object in the r82 image is certainly at least as compact
as any other object in the image. The R$-$K colour of $\sim$4.2 is
also redder than that of r73.

\subsection{Object r112}\label{sec:r112}

\begin{figure*}
 \begin{center}
  \leavevmode
  \epsfxsize 0.9\hsize
  \epsffile{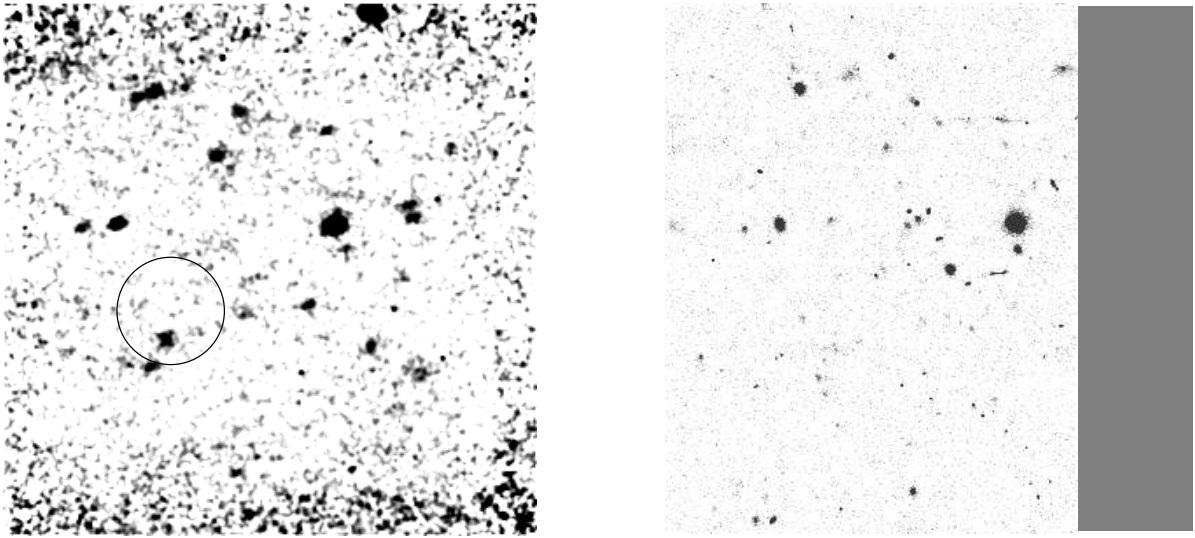}
 \end{center}
 \caption{Object r112. As figure~\protect\ref{fig:r73_piccies} but with slightly deeper K-band image (limiting K$\sim$20.5).
}\label{fig:r112_piccies}
\end{figure*}
Unlike the other sources, r112 has no single clear-cut candidate.  Instead,
several red objects, not visible in R, can be seen in the K-band image fairly
close to the X-ray centroid (a 10 arcsec error circle around the centroid is
shown in figure~\ref{fig:r112_piccies}). The details in
table~\ref{tab:candidates} refer to the object just inside the bottom of the
error circle, but the other nearby objects have similar colours (ie R$-$K$ \sim
4$). The images are too faint to determine whether or not the candidates are
significantly extended, but we are perhaps looking at a distant cluster (see
sec.~\ref{sec:clusters}).

\subsection{Object r130}\label{sec:r130}

\begin{figure*}
 \begin{center}
  \leavevmode
  \epsfxsize 0.9\hsize
  \epsffile{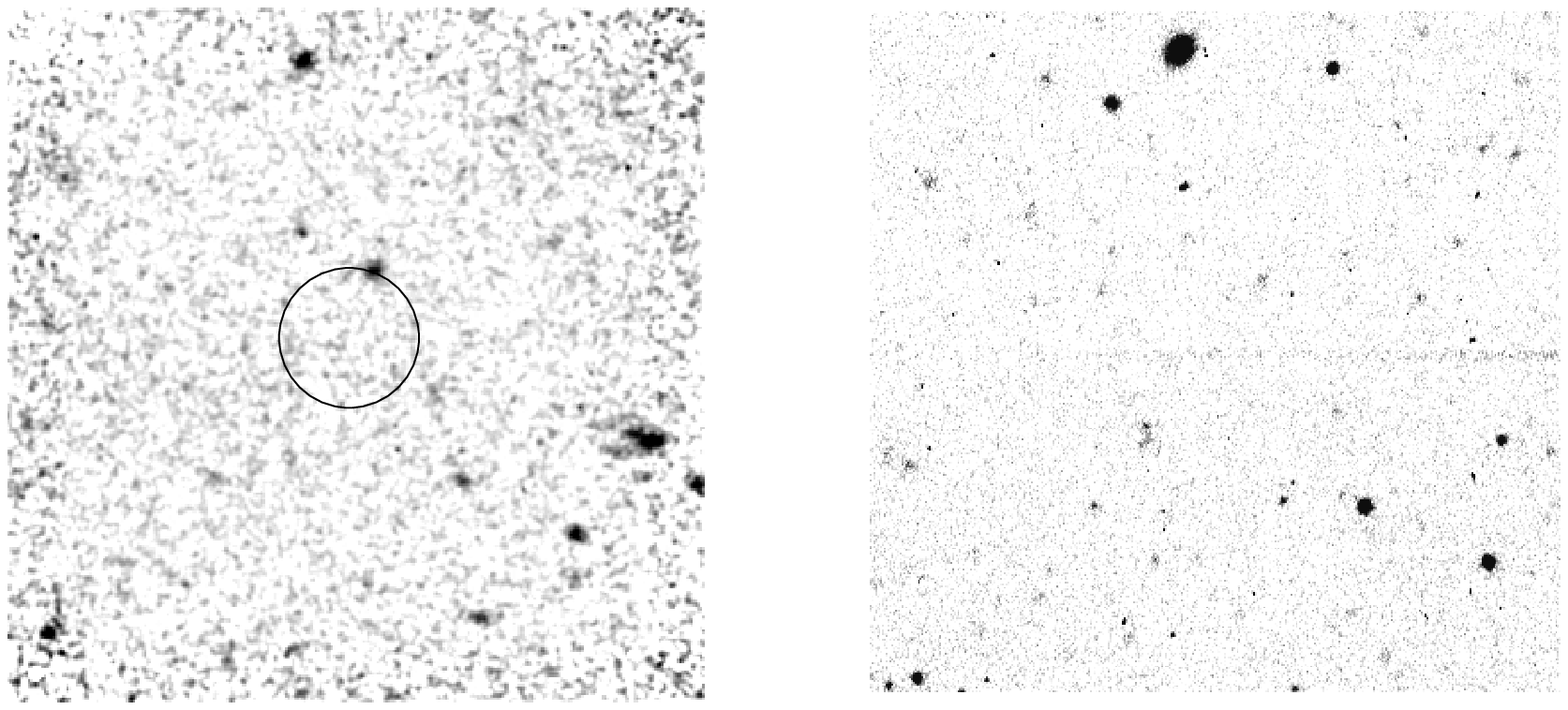}
 \end{center}
 \caption{Object r130. As figure~\protect\ref{fig:r73_piccies}.
}\label{fig:r130_piccies}
\end{figure*}
The final optically blank X-ray source, r130, is the reddest object of
all. There is very little to be found on the deep R-band image
(figure~\ref{fig:r130_piccies}), but a point-like object is seen on
the edge of the 10 arcsec error circle in the K-band image (10 arcsec
is quite a reasonable positional accuracy for an object this faint in
X-rays~---~see figure~\ref{fig:offsets}). The object is extremely
red~---~even though there is almost nothing to be seen in R, it is the
brightest candidate object found in any of our fields at K. It also
has an extremely high X-ray to optical flux ratio (see
table~\ref{tab:candidates} and figure~\ref{fig:fxfopt}).
%
\begin{table*}
\begin{tabular}{@{}c@{}cr@{$\;$}r@{$\;$}r@{$\;\;\;$}r@{$\;$}r@{$\;$}rcr@{$\;$}r@{$\;$}r@{$\;\;\;$}r@{$\;$}r@{$\;$}rc@{}c@{}}
\multicolumn{1}{c}{Object} &
 \multicolumn{1}{c}{{\rosat} flux} &
 \multicolumn{6}{c}{{\rosat} position} &
 \multicolumn{1}{c}{K mag.} &
 \multicolumn{6}{c}{Optical position} &
 \multicolumn{1}{c}{R$-$K} &
 \multicolumn{1}{c}{log[${\rm F}_{\rm X} / {\rm F}_{\hbox{opt}}$]} \\
\multicolumn{1}{c}{Name} &
 \multicolumn{1}{c}{(0.5~--~2 keV)} &
 \multicolumn{6}{c}{(J2000)} & 
 & 
 \multicolumn{6}{c}{(J2000)} &
 & \multicolumn{1}{c}{(R band)} \\
&
 \multicolumn{1}{c}{($\hbox{erg cm$^{-2}$ s$^{-1}$}$)} &
 \multicolumn{3}{c}{RA} & \multicolumn{3}{c}{Dec} & 
 & 
 \multicolumn{3}{c}{RA} & \multicolumn{3}{c}{Dec} &
 &\vspace{1mm} \\
  r73 & $4.13 \times 10^{-15}$ &  13&35&16.99 &37&54&18.9& 19.7 $\pm$ 0.25 & 13&35&17.2 &37&54&16$\phantom{.0}$& 3.2 $\pm$ 0.4 & 0.18 \\
  r82 & $3.85 \times 10^{-15}$ &  13&35&15.83 &37&52&41.1& 18.7 $\pm$ 0.25 & 13&35&15.9 &37&52&41.1&             4.2 $\pm$ 0.4 & 0.13 \\
 r112 & $2.81 \times 10^{-15}$ &  13&34&27.64 &37&54&22.7& 18.9 $\pm$ 0.25 & 13&34&27.5 &37&54&14$\phantom{.0}$& 4.2 $\pm$ 0.4 & 0.08 \\
 r130 & $2.25 \times 10^{-15}$ &  13&33&43.40 &37&50&32.1& 18.4 $\pm$ 0.25 & 13&33&43.1 &37&50&42.0&             6.1 $\pm$ 0.6 & 0.54 \\
\end{tabular}
\caption{X-ray, optical and IR data for the candidate objects for each `blank' 
 field as identified in sections~\ref{sec:r73} to~\ref{sec:r130}.
}\label{tab:candidates}
\end{table*}

\section{Possible Classifications}\label{sec:ident}

Although we cannot unambiguously classify the candidate objects with the
current data, we can narrow down the range of possibilities considerably.

\begin{figure}
 \epsfxsize \hsize
 \epsffile{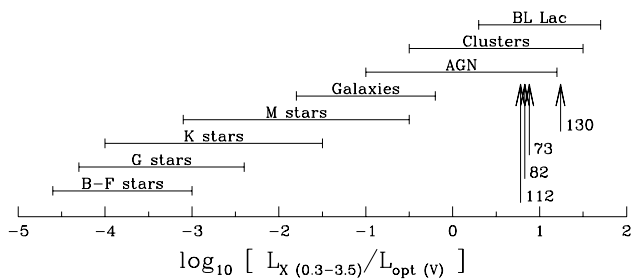}
 \caption{The horizontal bars show the ranges of {\lglxlopt} found by 
\protect\scite{Stocke+91} in the EMSS. For comparison, the {\lglxlopt} of the
blank fields are given. The optical flux for clusters is derived from the 
brightest cluster galaxy. The {\lxlopt} values for the blank fields have been 
converted from 0.5--2.0~keV to the 0.3--3.5~keV band used by Stocke assuming a 
spectral slope of 1.0 and converted from R to the V-band used by Stocke 
assuming a V$-$R of 1.
}\label{fig:lxlopt}
\end{figure}
\subsection{Stellar Sources}

In figure~\ref{fig:lxlopt} we compare the X-ray to optical luminosity ratios
({\lxlopt}) of the four candidates described above with the ratios found for
various classes of X-ray sources in the Einstein Medium Sensitivity Survey
(EMSS) by \scite{Stocke+91}. It is clear from this figure that a stellar
origin for the X-rays in any of the four blank fields can almost certainly be
ruled out. Although not shown in Stockes figure, it is perhaps possible that
we are looking at very faint cataclysmic variable (CV) systems, with the
K-band emission dominated by a faint M-dwarf companion. Such systems can give
very high {\lxlopt} values.

Although we do not have sufficient data to separate the contribution to the
measured brightness of the accretion disk and M-dwarf components,
\scite{Sproats+96} give a distance estimation technique for CVs which we can
use to get lower limits on the distances for the four candidates. To do this
we will assume that the M-dwarfs all have radii greater than $0.1$ solar
radii (see \pcite{Caillault+90}) and that the maximum V$-$R colour for the
objects is 4 \cite{Ramseyer94}. Under these assumptions we find that r130 is
$\ga 200$ pc away, r82 and r112 both $\ga 350$ pc and r73 $\ga 700$ pc. The
Deep Survey pointing is almost directly out of the galactic plane and
\scite{VanParadijs+96} find an exponential scale height for CVs of $\sim
200$pc. Therefore r130, r112 and r82 could all be within a reasonable
distance of the plane. Nevertheless, the rarity of comparable known objects
(eg \pcite{Kolb93}) means that this must be considered an intriguing but
largely unsupported possibility.

\subsection{Galaxies and NELGs}

In none of the K-band images are the likely candidate objects
significantly extended, however this is mainly due to the faintness of
the sources and the large IR background which makes modelling of the
profile difficult. Therefore, we cannot use apparent extension as a
means of identifying galaxies. The range of {\lxlopt} values for
``Normal Galaxies'' taken from by \scite{Stocke+91} in
figure~\ref{fig:lxlopt} is well away from the ratios found for the
blank fields, but this classification does not include the NELGs found
in the Deep Survey. As can be seen in figure~\ref{fig:fxfopt}, these
have X-ray to optical ratios that are similar to those of AGN and are,
therefore, much closer to the values for the blank field candidates
than normal galaxies. It is also important to remember that these
fields have, in essence, been {\em selected\/} because of their high
{\lxlopt} and it is not unreasonable to suppose that they will fall
towards the higher end of the distribution of {\lxlopt} for their
particular object type.

In addition, the R$-$K colour of object r73 of 3.2 is consistent with the NELGs
and, indeed, with most galaxies with $z \ll 1$ \cite{Steidel+95}. We have K
magnitudes for a handful of confirmed NELGs, and all have R$-$K between 3 and
3.5. However, the R$-$K of 4.2 for objects r82 and r112 is less reasonable and
R$-$K of $\sim 6$ for r130 is too extreme.

\subsection{Clusters of Galaxies}\label{sec:clusters}

The {\lxlopt} values for the blank field objects fall comfortably into the
range found for clusters of galaxies, but is the same true for the R-K
colours? If the X-ray emission from the objects is associated with a cluster,
then the candidate identifications will probably be the brightest cluster
galaxies (BCG). Under this assumption, we can use the apparent magnitude of
the object to estimate the redshift of the cluster and therefore check that
the R-K colours and absolute X-ray luminosity are reasonable.

\begin{table}
\begin{center}
\begin{tabular}{cccc}
Object & 
R-band & Redshift & $\log[{\rm L}_{\rm X}]$ \\
Name   & mag.  &   $z$& (0.5-2 keV)\\
       &
  & (estimate) & erg sec$^{-1}$\vspace{1mm}\\
    r73 & 22.9 & 0.8 & 43.0\\
    r82 & 22.9 & 0.8 & 43.1\\
   r112 & 23.1 & 0.9 & 42.9\\
   r130 & 24.5 & 1.1 & 43.1\\
\end{tabular}
\end{center}
\caption{Estimated redshifts and X-ray luminosities of the four X-ray sources
under the assumption that the candidate objects are the brightest members of a
galaxy cluster. Distance estimate calibration and K-corrections are taken from
\protect\scite{Schneider+83} with the K-corrections linearly extrapolated
slightly above $z=1$.}\label{tab:BCG}
\end{table}
Using the BCG calibration and K-corrections from \scite{Schneider+83} we
obtain the estimated redshifts and X-ray luminosities found in
table~\ref{tab:BCG}. Estimated redshifts for the four objects are between 0.8
and 1.1, with ${\rm L}_{\rm X}$ values all around $10^{43}$ erg
sec$^{-1}$. Other clusters and galaxy groups in the survey have $\log[{\rm
L}_{\rm X}]$ between 42 and 43.5 erg sec$^{-1}$ so all the candidates in the
blank fields are compatible with these. In addition, from \scite{Steidel+95}
we find that model galaxy spectra of elliptical galaxies give R-K colours of
$\sim$5 at $z=0.8$ up to $\sim$6 at $z=1.1$ which, given the scatter that is
seen around the model value, is again consistent with the R-K values found for
all the candidates (refer to table~\ref{tab:candidates}).

Therefore, a $z\sim 1$ galaxy cluster origin cannot be ruled out for any of the
blank fields.

\subsection{AGN}

Typically, AGN have fairly blue R$-$K colours of $\sim 2$, rarely going above 3
(see, for example, \pcite{Dunlop+86} figure~2). Therefore, the very red colours
of the candidate objects found, particularly of r82, r112 and r130, would seem
to exclude AGN as the source of the X-rays. However, such colours might be seen
in AGN under some circumstances.

For example, it is possible that this colour is a result of a very high
redshift. At $z>5$, Ly$_{\alpha}$ passes out of the R band, giving a very red
colour. This would make all four of the sources by far the most X-ray luminous
in the Deep Survey. While this is not impossible, it is unlikely to be the case
for all four sources and so, perhaps more realistically, we should consider the
possibility of intrinsic reddening. It has frequently been suggested that
obscured AGN could contribute a significant part of the XRB at a large range of
energies, not just the soft X-ray band (eg~\pcite{Madau+94};
\pcite{Comastri+95}). The recent discovery by \ncite{Shanks+96}
(\ycite{Shanks+96}; see also \pcite{Almaini+95}) of an X-ray source with narrow
optical emission lines, but a broad IR ${\rm H}_{\alpha}$ has provided a
probable member of this class of object. The object ({\rosat} object
RXJ13434+0001 is also reddened with R$-$K$\sim 4.5$ and an ${\rm F}_{\rm X} /
{\rm F}_{\hbox{opt}}$ of $\sim$ 0.45 in the same units as
table~\ref{tab:candidates}). This is similar to object r82 and r130 could be an
even more highly obscured AGN.

The amount of reddening required to give the observed colour will depend to
some extent upon redshift.  For example, to produce an observed R$-$K colour
of 6 for an object with an intrinsic colour of 2 requires an $N_H$ column of
$\sim 10^{21.8} \hbox{cm}^{-2}$ at a redshift of $z=1$. (For this and
subsequent calculations we assume a dust to gas ratio of $5.8 \times
10^{21}\;\hbox{cm}^{-2}\ \hbox{mag}^{-1}$ taken from \pcite{Bohlin+78}). This
would take the intrinsic {\lglxlopt} of r130 down to $\sim -3$~---~far too low
for an extra-galactic object. On the other hand, the same observed colour for
an object at $z=2$ would need $N_H \sim 10^{21.5} \hbox{cm}^{-2}$ which gives
an intrinsic {\lglxlopt} for r130 of only -1, which although low is certainly
acceptable for an AGN (see figure~\ref{fig:lxlopt}).

For objects r82 and r112 a lower column is required to produce the observed
R$-$K colours. A distance of $1\la z \la 3$ would be reasonable to produce an
intrinsic {\lxlopt} suitable for an AGN with an $N_H$ of between $\sim
10^{21}$ and $10^{21.5}\;\hbox{cm}^{-2}$.

Alternatively, we could be looking at BL Lacs. These have redder colours than
other AGN, typically in the range $2\la {\hbox{R$-$K}} \la 4.5$ (eg
\pcite{Cruz-Gonzalez+83}) and high {\lxlopt} values (see
figure~\ref{fig:lxlopt}). Therefore, only the R$-$K of 6.1 for r130 is too
red. However, BL Lacs should be detectable in the radio, but deep 20cm and 6cm
radio observations of the Deep Survey region at the Very Large Array (VLA)
show no signs of radio sources down to the detection limit ($\sim 0.3$ mJy)
for any of the four blank fields.

\section{Conclusions}\label{sec:conclusions}

We have used deep K-band imaging to attempt to identify the optical candidates
of four faint X-ray sources with `blank' optical fields (ie nothing within
$\sim 15$ arcsec of the X-ray centroid with R$\la$23). Of the four fields,
firm candidates have been found in three. The remaining field (object r112;
figure~\ref{fig:r112_piccies}) shows a collection of similar, red objects in
the vicinity of the X-ray centroid. No firm conclusions about the nature of
the candidates can be achieved with just the current photometric data but we
have, nevertheless, been able to draw tentative conclusions.

Of the four X-ray sources, r73 has the least red candidate. In fact, both its
R$-$K colour (3.2) and X-ray to optical luminosity ratio are entirely
consistent with the Narrow Emission Line Galaxies (NELGs) found in significant
quantities at the fainter end of the UK {\rosat} Deep Survey \cite{mn_nov}.

On the other hand, both r112 and r130 have candidates that are redder than in
r73 (r112 has an R$-$K of 4.2, r130 has R$-$K of 6.1). These objects are
probably better described by high redshift ($z\sim 1$) galaxy clusters, the
candidates having absolute magnitudes, R$-$K colours and {\lxlopt} ratios
consistent with typical $z\sim 1$ brightest cluster galaxys (BCGs). The
collection of red objects in r112 also supports this conclusion.

The candidate object for source r82 also is also consistent with being a
distant brightest cluster galaxy. However, in this case both the deep R and
K-band images show an apparently point-like object. It may therefore also be
explained as a distant ($z\sim 2$) intrinsically reddened AGN. If this is the
case, with this deep {\rosat} observation, and others like it, we are just
starting to see the tip of the iceberg which has been proposed as a significant
contributor to the cosmic X-ray background, not just in the soft band, but over
a wide range of energies (eg~\pcite{Madau+94}). 

Future observations will be required to confirm these tentative
identifications. In particular, infra-red spectroscopy over a range of
wavelengths would enable one to search for broad, redshifted emission lines
characteristic of QSOs (a QSO with a redshift of $z\sim 2$ would have H$\alpha$
redshifted into the H or K band). Further IR photometry in other bands to give
IR colours may also prove a useful diagnostic and deeper imaging with good
seeing would allow spatial extension to be seen if any of the objects are NELGs
or moderate redshift clusters (ie~$z\ll 1$).

Deep IR imaging, therefore, has shown itself to be a powerful tool in the
study of the faintest X-ray sources. In all four of the optically blank fields
containing X-ray sources which we have observed in the K-band, firm candidates
have been found. In addition, consideration of the photometric and X-ray
properties of the candidates has enabled us to make preliminary
identifications of the types of objects producing the X-ray emission, and
possibly identified an obscured QSO~---~a much hypothesised but little
observed contributor to the cosmic X-ray background.

\section*{Acknowledgments}

Thanks to Tom Marsh for discussions about possible galactic identifications and
to the referee for his encouraging comments.

The United Kingdom Infrared Telescope (UKIRT) is operated by the Joint
Astronomy Centre on behalf of the U.K. Particle Physics and Astronomy Research
Council. 

IRAF is distributed by the National Optical Astronomy Observatories, which are
operated by the Association of Universities for Research in Astronomy, Inc.,
under cooperative agreement with the National Science Foundation.


\begin{thebibliography}{99}
\bibitem[\protect\citename{Almaini {\etal} }{1995}]{Almaini+95}
 Almaini, O., Boyle, B.J., Griffiths, R.E., Shanks, T., Stewart, G.C., 
 Georgantopoulos, I., 1995, MNRAS, 277, L31.
\bibitem[\protect\citename{Almaini {\etal} }{1996}]{Almaini+96}
 Almaini, O., Shanks, T., Boyle, B. J., Griffiths, R. E.,
 Roche, N.; Stewart, G. C., Georgantopoulos, I., 1996, MNRAS, 282, 295.
\bibitem[\protect\citename{Bessell }{1991}]{Bessell91}
 Bessell, M.S., 1991, Astron. J., 101, 662.
\bibitem[\protect\citename{Bohlin, Savage \& Drake }{1978}]{Bohlin+78}
 Bohlin, R.C., Savage, B.D., Drake, J.F., 1978, ApJ, 224, 132.
\bibitem[\protect\citename{Boyle {\etal} }{1994}]{Boyle+94}
 Boyle, B.J., Shanks, T.,  Georgantopoulos, I., Stewart, G.C.,
 Griffiths, R.E., 1994, MNRAS, 271, 639.
\bibitem[\protect\citename{Boyle {\etal} }{1995}]{Boyle+95}
 Boyle, B.J, M$^{\rm c}$Mahon, R.G., Wilkes, B.J., Elvis, M.,
 1995, MNRAS, 276, 315.
\bibitem[\protect\citename{Branduardi-Raymont {\etal} }{1994}]{GBR}
 Branduardi-Raymont, G., et al., 1994, MNRAS, 270, 947.
\bibitem[\protect\citename{Caillault \& Patterson }{1990}]{Caillault+90}
 Caillault, J-P., Patterson, J., 1990, Astron. J., 100, 825.
\bibitem[\protect\citename{Cash }{1979}]{Cash79}
 Cash, W., 1979, ApJ, 228, 939.
\bibitem[\protect\citename{Comastri {\etal} }{1995}]{Comastri+95}
 Comastri, A., Setti, G., Zamorani, G., Hasinger, G., 1995, A\&A, 296, 1.
\bibitem[\protect\citename{Cruz-Gonzalez \& Huchra }{1983}]{Cruz-Gonzalez+83}
 Cruz-Gonzalez, I., Huchra, J.P., 1983, ApJ, 89, 441.
\bibitem[\protect\citename{Dunlop {\etal} }{1986}]{Dunlop+86}
 Dunlop, J.S., Downes, A.J.B., Peacock, J.A., Savage, A., Lilly, S.J., Watson, 
 F.G., Longair, M.S., 1986, Nature, 319, 564.
\bibitem[\protect\citename{Jones {\etal} }{1995}]{Jones+94/5}
 Jones, L.R. {\etal} 1995. Proc 35th Herstmonceux Conf., ed Mattox, S.
\bibitem[\protect\citename{Jones {\etal} }{1996}]{Jones+96}
 Jones, L.R., M$^{\rm c}$Hardy, I.M., Merrifield, M.R., Mason, K.O., 
 Smith, P.J., Branduardi-Raymont, G., Newsam, A.M., Dalton, G., 
 Rowan-Robinson, M., Luppino, G., 1996, MNRAS (in press).
\bibitem[\protect\citename{Kolb }{1993}]{Kolb93}
 Kolb, U., 1993, A\&A, 271, 149.
\bibitem[\protect\citename{Madau {\etal} }{1994}]{Madau+94}
 Madau, P., Ghisellini, G., Fabian, A.C., 1994, MNRAS, 270, L17
\bibitem[\protect\citename{M$^{\rm c}$Hardy {\etal} }{1995}]{McHardy+95/6}
 M$^{\rm c}$Hardy, I.M. {\etal}, 1995, MPE Report 263, Proceedings of 
 `R\"ontgenstrahlung from the Universe', p331.
\bibitem[\protect\citename{M$^{\rm c}$Hardy {\etal} }{1997}]{mn_nov}
 M$^{\rm c}$Hardy, I.M. {\etal}, 1997, MNRAS (submitted).
\bibitem[\protect\citename{Metzger, Luppino \& Miyazaki }{1995}]{8kpaper}
 Metzger, M.R., Luppino, G.A., Miyazaki, S., 1995, BAAS, 187, 73.
\bibitem[\protect\citename{Van Paradijs, Augusteijn \& Stehle }{1996}]%
{VanParadijs+96}
 Van Paradijs, J., Augusteijn, T., Stehle, R., 1996, A\&A, 312, 93.
\bibitem[\protect\citename{Ramseyer }{1994}]{Ramseyer94}
 Ramseyer, T.F., 1994, ApJ, 425, 243.
\bibitem[\protect\citename{Romero-Colmenero {\etal} }{1996}]{R-C+96}
 Romero-Colmenero, E., Branduardi-Raymont, G., Carrera, F.J.,
 Jones, J.R., Mason, K.O., M$^{\rm c}$Hardy, I.M., Mittaz, J.P.D., 1996,
 MNRAS, 282, 94.
\bibitem[\protect\citename{Schneider, Gunn \& Hoessel }{1983}]{Schneider+83}
 Schneider, D.P., Gunn, J.E., Hoessel, J.G., 1983, ApJ, 264, 337.
\bibitem[\protect\citename{Shanks {\etal} }{1991}]{Shanks+91} Shanks, T., 
 Georgantopoulos, I., Stewart, G.C., Pounds,
 K.A., Boyle, B.J., Griffiths, R.E., 1991, Nature, 353, 315.
\bibitem[\protect\citename{Shanks {\etal} }{1996}]{Shanks+96}
 Shanks, T., Almaini, O., Boyle, B.J., Della-Ceca, R., Done, C.,
 Georgantopoulos, I., Griffiths, R.E., Rawlings, S.J., Roche, N., Stewart, G.C.,
 MPE Report 263, Proceedings of `R\"ontgenstrahlung from the Universe', p341.
\bibitem[\protect\citename{Sproats, Howell \& Mason }{1996}]{Sproats+96}
 Sproats, L.N., Howell, S.B., Mason, K.O., 1996, MNRAS, 282, 1211.
\bibitem[\protect\citename{Steidel \& Dickinson }{1995}]{Steidel+95}
 Steidel, C.C., Dickinson, M., in `Wide Field Spectroscopy and the Distant 
 Universe', 1995, p349.
\bibitem[\protect\citename{Stocke {\etal} }{1991}]{Stocke+91}
 Stocke, J.T., Morris, S.L., Gioia, I.M., Maccacaro, T., Schild, R.,
 Wolter, A., Fleming, T.A., Henry, J.P., 1991, ApJS, 76, 813.
\end{thebibliography}
\end{document}